\documentclass[conference]{IEEEtran}
\IEEEoverridecommandlockouts
\usepackage{amsmath,amsfonts}
\usepackage{algorithm}
\usepackage{algpseudocode}
\usepackage{amsmath}
\usepackage{array}
\usepackage[caption=false,font=footnotesize,labelfont=rm,textfont=rm]{subfig}
\usepackage{textcomp}
\usepackage{stfloats}
\usepackage{url}
\usepackage{verbatim}
\usepackage{graphicx}
\usepackage{subfloat}
\usepackage{float}
\usepackage{colortbl}

\usepackage{booktabs}  
\usepackage{threeparttable} 
\usepackage{multirow}
\usepackage{makecell}
\usepackage{diagbox}
\usepackage{pifont}

\usepackage[font=footnotesize]{caption}

\usepackage{cite}

\def\BibTeX{{\rm B\kern-.05em{\sc i\kern-.025em b}\kern-.08em
    T\kern-.1667em\lower.7ex\hbox{E}\kern-.125emX}}
\begin{document}

\title{FsimNNs: An Open-Source Graph Neural Network Platform for SEU Simulation-based Fault Injection
}

\author{\IEEEauthorblockN{Li Lu\IEEEauthorrefmark{1,}, Jianan Wen\IEEEauthorrefmark{1,}\IEEEauthorrefmark{2}, Milos Krstic\IEEEauthorrefmark{1,}\IEEEauthorrefmark{2}}
\IEEEauthorblockA{\IEEEauthorrefmark{1}IHP-Leibniz-Institut für innovative Mikroelektronik, Frankfurt (Oder), Germany}
\IEEEauthorblockA{\IEEEauthorrefmark{2}University of Potsdam, Potsdam, Germany}
{Emails: \{lu, wen, krstic\}@ihp-microelectronics.com}
}

\maketitle

\begin{abstract}
Simulation-based fault injection is a widely adopted methodology for assessing circuit vulnerability to Single Event Upsets (SEUs); however, its computational cost grows significantly with circuit complexity. To address this limitation, this work introduces an open-source platform that exploits Spatio-Temporal Graph Neural Networks (STGNNs) to accelerate SEU fault simulation. The platform includes three STGNN architectures incorporating advanced components such as Atrous Spatial Pyramid Pooling (ASPP) and attention mechanisms, thereby improving spatio-temporal feature extraction. In addition, SEU fault simulation datasets are constructed from six open-source circuits with varying levels of complexity, providing a comprehensive benchmark for performance evaluation. The predictive capability of the STGNN models is analyzed and compared on these datasets. Moreover, to further investigate the efficiency of the approach, we evaluate the predictive capability of STGNNs across multiple test cases and discuss their generalization capability. The developed platform and datasets are released as open-source to support reproducibility and further research on https://github.com/luli2021/FsimNNs.

\end{abstract}

\begin{IEEEkeywords}
Simulation-based fault injection, SEU faults, Machine Learning, Graph Neural Networks 
\end{IEEEkeywords}

\section{Introduction}
As integrated circuits scale to nanoscale dimensions, their susceptibility to faults from external disturbances increases. Single Event Upsets (SEUs) \cite{SEU}, a prevalent form of soft error, are particularly significant in high radiation environments such as space, but may also occur at ground level due to cosmic rays and other ionizing sources \cite{SEEsource1}. Consequently, SEU-related reliability analysis has become essential for applications in space, automotive, and high-performance computing systems. 

SEUs typically manifest as transient bit flips in storage elements such as flip-flops, with errors persisting until overwritten. Despite their transient nature, SEUs can cause data corruption or critical system failures. Simulation-based fault injection (i.e., fault simulation) \cite{FI} is commonly used to evaluate SEU impacts during early design stages. However, as circuit complexity increases, simulation time grows exponentially. Various techniques have been proposed to address this challenge, each with its limitations \cite{STGCN_TCAD}. For example, statistical fault injection (SFI) \cite{SFI1} reduces simulation time by sampling a subset of possible faults. However, its effectiveness on complex circuits cannot be guaranteed, as the fault sample representation may be inadequate. Using an inappropriate sampling method could lead to significant discrepancies between simulation and actual outcomes \cite{SFIDis1}.

Machine Learning (ML) techniques have been increasingly adopted in recent years to automate processes and reduce computational costs across a variety of domains. In the context of reliability analysis, our work in \cite{STGCN_TCAD} demonstrated the use of a Spatio-temporal Graph Convolutional Network (STGCN) to predict Single Event Upset (SEU) fault simulation results. STGCN \cite{STGCN}, a representative architecture of Spatio-Temporal Graph Neural Networks (STGNNs), belongs to the broader class of Graph Neural Networks (GNNs) \cite{GNNsurvey}, which are designed to model graph-structured data. By representing digital circuits as graphs, GNNs can effectively capture gate-level connectivity, while STGNNs extend this capability by incorporating temporal dependencies from workload simulations. Building upon these advances, this paper presents a more comprehensive investigation of the STGCN approach with respect to both STGNN architectural design and application scenarios. Furthermore, an open-source platform, referred to as FsimNNs, has been developed to support and facilitate this approach.

\begin{figure}[!t]
\centering
\includegraphics[width=0.46\textwidth]{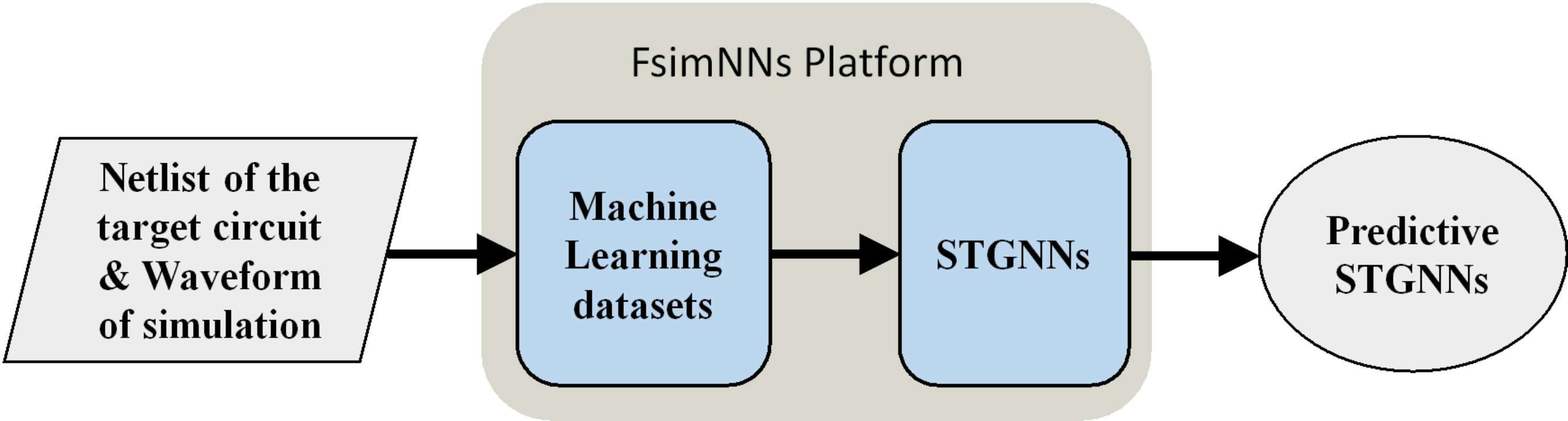}
\caption{Overview of FsimNNs within the STGNN approach framework.}
\label{FsimNNs}
\vspace{-0.5cm}
\end{figure}

The overall framework of the STGNN approach is shown in Fig. \ref{FsimNNs}. Specifically, features related to SEU fault simulation are extracted from the target circuit’s netlist and corresponding simulation waveforms to build the ML datasets. Using these datasets, STGNN models are trained, and the optimal model is selected to predict the outcomes of SEU fault simulations. Our open-source platform provides both the datasets generated on six circuits (see Chapter \ref{MLDataGen}) and the STGNN models (see Chapter \ref{STGNNModels}). For researchers wanting to apply this method to their own circuits, the dataset can be generated using the procedure described in Chapter\ref{FeatureExtraction}. The main contributions of this paper are summarized as follows:

•	To expand the selection of STGNN models, two new architectures are introduced, incorporating advanced components like Atrous Spatial Pyramid Pooling (ASPP) and attention mechanisms. To offer references for new applications, their performance is systematically assessed and compared with that of STGCN using datasets from six open-source circuits with varying functionality and complexity.

•	The potential of STGNNs to accelerate SEU fault simulation is further explored by evaluating their ability to generalize across test cases. This evaluation is conducted on Ibex, an open-source RISC-V core.

•	To promote reproducibility and support further research, the proposed STGNN models and ML datasets of SEU fault simulation are released as open-source at https://github.com/luli2021/FsimNNs.

\section{Datasets}
\label{MLDataGen}
This section details the datasets provided by the FsimNNs platform and the process for generating them from raw SEU fault simulation of a circuit, enabling researchers to construct custom datasets.

\renewcommand{\arraystretch}{1.2}
\begin{table}[t]
     \setlength\tabcolsep{2.5pt} 
	\centering
	\fontsize{8}{10}\selectfont
	\begin{threeparttable}
		\caption{Test Circuits. \textit{Num\_ff} indicates the number of flip-flops in the circuit; \textit{Num\_time} indicates the number of SEU fault injection time points; the total number of SEU faults (\textit{Num\_total}) is calculated as the product of \textit{Num\_ff} and \textit{Num\_time}. Take RI5CY as an example, there are 121,640 samples ($3041 \times 40$).}
		\label{tab:datasets}
		\begin{tabular}{ccccccccccc}
			\toprule[1.1pt]
			{}&{SPI}&{I2C}&{XGE-MAC}&{ETH-MAC}&{Ibex}&{RI5CY}\cr 
			\midrule
			{Num\_ff}&{131}&{153}&{1358}&{1233}&{2126}&{3041}\cr
			{Num\_time}&{50}&{50}&{36}&{32}&{46}&{40}\cr
			{Num\_total}&{6550}&{7650}&{48888}&{39456}&{97796}&{121640}\cr
			\hline
			\bottomrule[1.5pt]
		\end{tabular}
	\end{threeparttable}
\vspace{-0.5cm}
\end{table}

\subsection{Raw data of SEU fault simulation}
\label{testCircuits}
Table \ref{tab:datasets} presents the details of the datasets provided by the platform to validate the STGNN approach. The datasets are built on six circuits exhibiting varying functionalities and complexities. Four of them, including I2C, SPI, XGE-MAC, and ETH-MAC, are obtained from the OpenCores \cite{OpenCores}, which provides Register-Transfer Level (RTL) IPs and corresponding test benches. The remaining two circuits are well-known open-source RISC-V cores: RI5CY \cite{RI5CY} and Ibex \cite{IBEX}. Ibex, a 32-bit RISC-V CPU core, is designed for embedded control applications and has undergone extensive verification with a provided UVM test bench \cite{IBexUVM}. For RI5CY, a more straightforward test bench is provided by the PULPissimo platform \cite{PULP}.

Our reference simulations are conducted using the Xcelium fault simulator \cite{xrun} provided by Cadence. In the simulation, primary outputs are used as observation points. Each SEU fault is injected individually, and if the output of the fault simulation differs from the fault-free simulation, the fault is classified as detected. If no valid output difference is observed, it is marked as undetected. All circuits are synthesized to gate-level netlists, and simulations are implemented at the gate level.

In our STGNN approach, SEU fault simulation characteristics are transformed into spatio-temporal graphs to meet the model's input requirements. The objective is to classify each node, representing an SEU fault sample, as either positive (detected) or negative (undetected). The following subsection details the graph generation method.

\begin{figure*}[!t]
\centering
\includegraphics[width=0.98\textwidth]{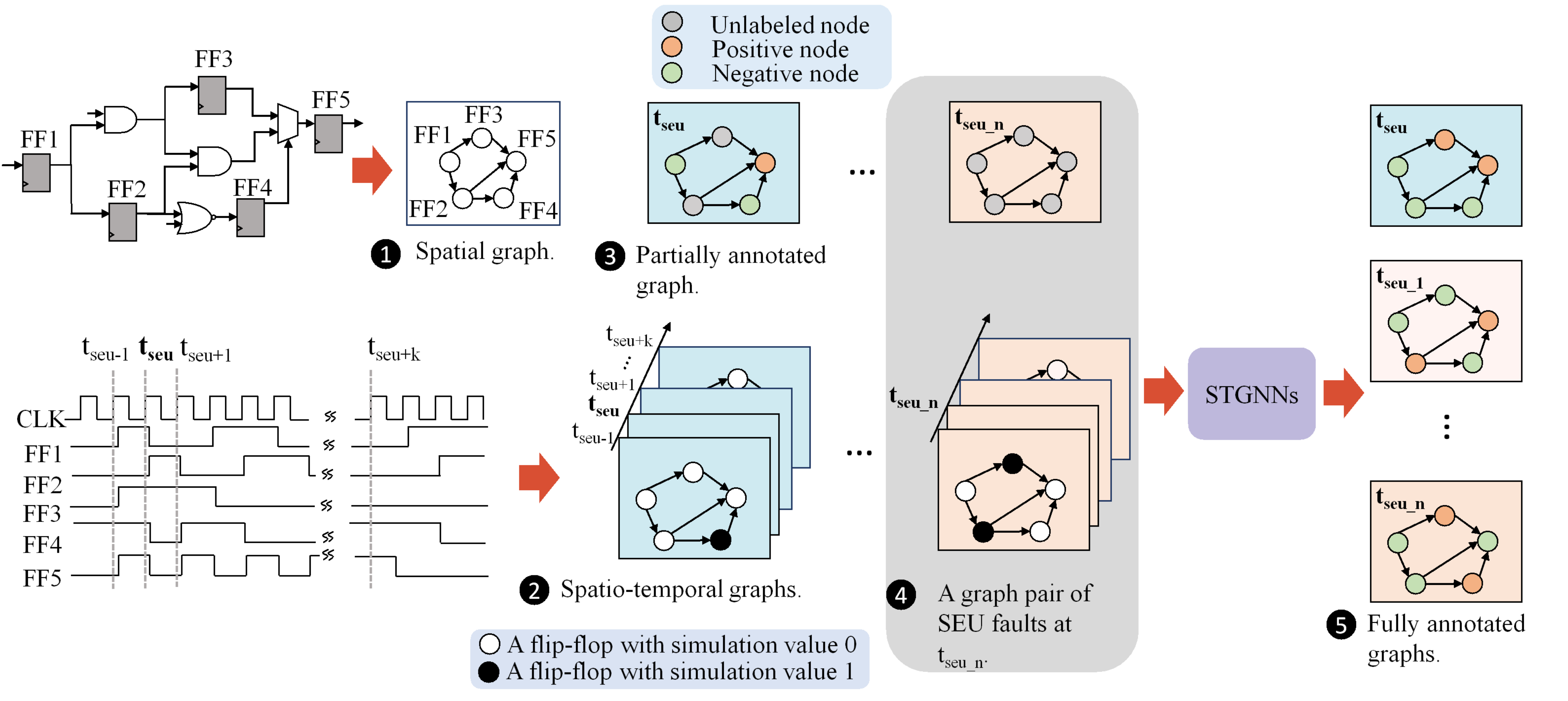}
\caption{The process of feature extraction and STGNN training.}
\label{overview}
\vspace{-0.5cm}
\end{figure*}

\subsection{Spatio-temporal graph generation}
\label{FeatureExtraction}
We use the same method proposed in the study \cite{STGCN_TCAD} to generate spatio-temporal graphs. The process is illustrated in Fig. \ref{overview}. Each graph is denoted as $G = (V, E)$, where $V$ and $E$  represent the sets of nodes and edges, respectively. The circuit under simulation is described using 1) a feature tensor $X \in R^{t \times n \times m}$, where $t$ is spatio-temporal graphs' time window size, $n$ is the number of nodes, and $m$ is the dimension of each node's feature vector; 2) an adjacency matrix $A \in R^{n \times n}$ indicating node connectivity, where $A_{ij} = 1$ when there is a connection between node $i$ and $j$, otherwise $A_{ij} = 0$; 3) an edge tensor $Y \in R^{n \times n \times c}$ capturing edge attributes, where $c$ denotes the dimension of the edge feature vectors.

In the first step, spatial features are extracted from the target circuit's netlist to construct spatial graphs (see \ding{182}), represented by the adjacency matrix $A$ and edge tensor $Y$. Each flip-flop is modeled as a node, and Breadth-First Search (BFS) is used to identify the shortest path between flip-flops. The distance between two flip-flops is defined by the number of combinational gates along the shortest path. An edge is established between two flip-flops if the distance is less than or equal to a predefined threshold, $max\_distance$. To exploit the characteristics of combinational gates on the paths, edge vectors are used to represent the type and order of the combinational gates on the edges \cite{STGCN_TCAD}.

In the second step, temporal features are incorporated into the graph structure by generating the feature tensor $X$ based on the Value Change Dump (VCD) waveform of the simulation test. As an example, to analyze SEU faults injected at a specific time $t_{seu}$, spatio-temporal graphs (see \ding{183}) are constructed to capture the temporal behavior around the time. The time window size, denoted as $time\_win\_size$, defines the number of clock cycles included in the graphs. The spatio-temporal graph at time $t_{seu-1}$ (a cycle previous of $t_{seu}$) is constructed using the simulation value (see the first graph in \ding{183}), where the value of \textit{FF4} is 1 and all other flip-flops have a value of 0. Following the same procedure, spatio-temporal graphs for clock cycles $t_{seu}$, ..., $t_{seu+k}$ (k cycles after $t_{seu}$) are generated based on the simulation values of flip-flops at their respective time points. The resulting feature tensor $X$ ($X \in R^{ t\times n \times m}, t=time\_win\_size=k+2$, $m=1$, $n$ is the number of nodes) representing the spatio-temporal features of all flip-flops around the time $t_{seu}$ is generated. The spatio-temporal graphs for other injection time points are generated using the same method.

\subsection{Sampling and training}
An effective sampling strategy for selecting annotated samples to train STGNNs is critical for achieving high prediction performance. Graph \ding{184} contains labels indicating the simulation results of SEU faults injected at a specific time point ($t_{seu}$), where a subset of nodes is annotated and the remaining nodes are left unlabeled. For each injection time point from $t_{seu}$ to $t_{seu\_n}$, the graph is paired with the corresponding spatio-temporal graphs generated in the previous subsection to form a graph pair (see \ding {185}, the pair is for time $t_{seu\_n}$). STGNNs are trained on these graph pairs to learn the mapping between input features and labels, thereby acquiring prediction capabilities. Ultimately, STGNNs can output fully annotated graphs (see \ding {186}) indicating the simulation results for all SEU samples.

The work \cite{STGCN_TCAD} introduced three sampling methods: spatial, temporal, and hybrid. Spatial sampling randomly selects flip-flops at each fault injection time and annotates the corresponding nodes, while temporal sampling selects fault injection time points. This work adopts hybrid sampling across both spatial and temporal domains.

\section{STGNN models}
\label{STGNNModels}
This section describes the architectures of the STGNN models used in this work. Although several advanced STGNNs have been proposed in earlier studies, they are not directly suitable for our problem. For instance, Graph WaveNet \cite{GraphWavenet} uses an adaptive adjacency matrix learned from data, removing the need for prior knowledge of the graph structure. However, this approach is not appropriate for our application, where a predefined circuit topology is essential. Therefore, we design customized STGNN architectures by integrating advanced components, as explained in the following subsections, to improve performance and offer more options for future research.

\subsection{Spatial feature learning}
To learn spatial dependencies representing the circuits' structural features in the data, we use two components: the Graph Convolutional Layer and the Graph Attention Layer, to build various STGNNs, which allows us to choose the best one for a specific dataset.

\subsubsection{Graph Convolutional Layer}
\label{GCL}
The Graph Convolutional Layer, introduced in GCNs \cite{GCN}, can be interpreted as learning a set of aggregation functions that integrate the features of a node's neighbors to update the node's representation. The node-wise formulation for computing the feature of a central node is expressed as follows:
\begin{align}
s_{ji} &= \mathbf{y}_{ji} \Theta_e ^{T} \label {eq:GCN1} \\
\mathbf{x'}_i &= \sigma\left( \Theta ^{T} \sum_{j \in N(i) \cup i} \frac {s_{ji}} {c_{ji}} \mathbf{x}_j \right) \label {eq:GCN2}
\end{align}
In \eqref{eq:GCN1}, $\mathbf{y}_{ji}$ represents the edge vector from node $j$ to node $i$, while $\Theta_e ^{T}$ denotes the transposition of the weight matrix to transform $\mathbf{y}_{ji}$ to the scalar weight $s_{ji}$. Equation \eqref{eq:GCN2} calculates the feature of node $i$ ($\mathbf{x'}_i$) based on the features of its neighbors and itself, where $N(i)$ signifies the set of neighbors of node $i$, $\mathbf{x}_j$ indicates the feature of node $j$, $c_{ji}$ is the product of the square roots of the degrees of nodes $i$ and $j$, $\sigma$ is the activation function, and $\Theta^ {T}$ represents the transposition of the weight matrix.

\subsubsection{Graph Attention Layer}
The attention mechanism is a widely used technique in ML applications. Graph Attention Networks (GAT) \cite{GAT} introduces attention mechanisms in GNNs. In GCN, the contributions of neighboring nodes to the central node are determined by $c_{ji}$ and the edge weight, as shown in \eqref{eq:GCN2}. In contrast, GAT employs the attention mechanism to learn the relative weights between two connected nodes. Its node-wise formulation is given by \eqref{eq:GAT1}-\eqref{eq:GAT3}. 
\begin{align}
\mathbf{e}_{ij} &= \mathbf{y}_{ij} \Theta_e ^{T}  \label {eq:GAT1} \\
\alpha_{ij} &= \frac {\exp( \phi( \mathbf{a}_s^T \Theta \mathbf{x}_i  +  \mathbf{a}_t^T \Theta \mathbf{x}_j  + \mathbf{a}_e^T \Theta_e \mathbf{e}_{ij} ) )} { \sum_{k \in N(i) \cup i} \exp( \phi( \mathbf{a}_s^T \Theta \mathbf{x}_i  +  \mathbf{a}_t^T \Theta \mathbf{x}_k +  \mathbf{a}_e^T \Theta_e \mathbf{e}_{ik} ))} \label {eq:GAT2} \\
\mathbf{x'}_i &= \sigma\left( \sum_{j \in N(i) \cup i} \alpha_{ij} \Theta \mathbf{x}_j \right) \label {eq:GAT3}
\end{align}
The equation \eqref{eq:GAT1} represents the linear transformation to calculate the $\mathbf{e}_{ij}$, the edge vector from node $i$ to node $j$. The equation \eqref{eq:GAT2} calculates the attention coefficient $\alpha_{ij}$, which indicates the degree of attention that node $i$ pays to node $j$. Here, $\phi$ denotes the LeakyReLU activation function, while $\mathbf{a}_s$, $\mathbf{a}_t$, and $\mathbf{a}_e$ are the attention weight vectors for the source node, target node, and edge, respectively.

\subsection{Temporal feature learning}
We must incorporate components to capture the temporal dynamic behaviors of the circuit executing a workload. As discussed in STGCN\cite{STGCN}, previous works employ Recurrent Neural Network (RNN) components to learn along the time axis. However, RNN-based networks can be challenging to train and demand significant computational resources. Therefore, STGCN utilizes a fully convolutional structure for temporal feature learning. We adopt similar strategies in this work by incorporating two types of components. 

\subsubsection{Temporal Convolutional Layer}
Atrous convolution \cite{DCov} is employed to implement the Temporal Convolutional Layer. This convolution enables the extraction of multi-scale features with fewer parameters. Our work applies it along the temporal dimension with a kernel size of 2 and varying dilation rates. The output of a one-dimensional atrous convolution is defined as follows:
\begin{align}
\mathbf{y}[i] &= \sum_{k=1}^{K} \mathbf{x}[i + r \cdot k]  \cdot \Theta[k] \label {eq:Atrous}
\end{align}
where $\mathbf{x}$ represents the input vector, $\mathbf{y}$ denotes the output vector, $\Theta$ signifies the weight of the convolution filter, $r$ indicates the dilation rate, $K$ refers to the filter size (which is 2 in our case).

\subsubsection{Atrous Spatial Pyramid Pooling (ASPP)}
Atrous Spatial Pyramid Pooling (ASPP), introduced in DeepLabv2 \cite{DeepLabv2}, applies multiple parallel atrous convolutions with varying dilation rates. This technique enhances the model’s ability to recognize objects at different scales. The ASPP implemented in this work is based on the Temporal Convolutional Layer above. It enables STGNNs to capture both fine-grained details and global temporal information in the data.

\begin{figure}[!t]
\centering
\includegraphics[width=0.48\textwidth]{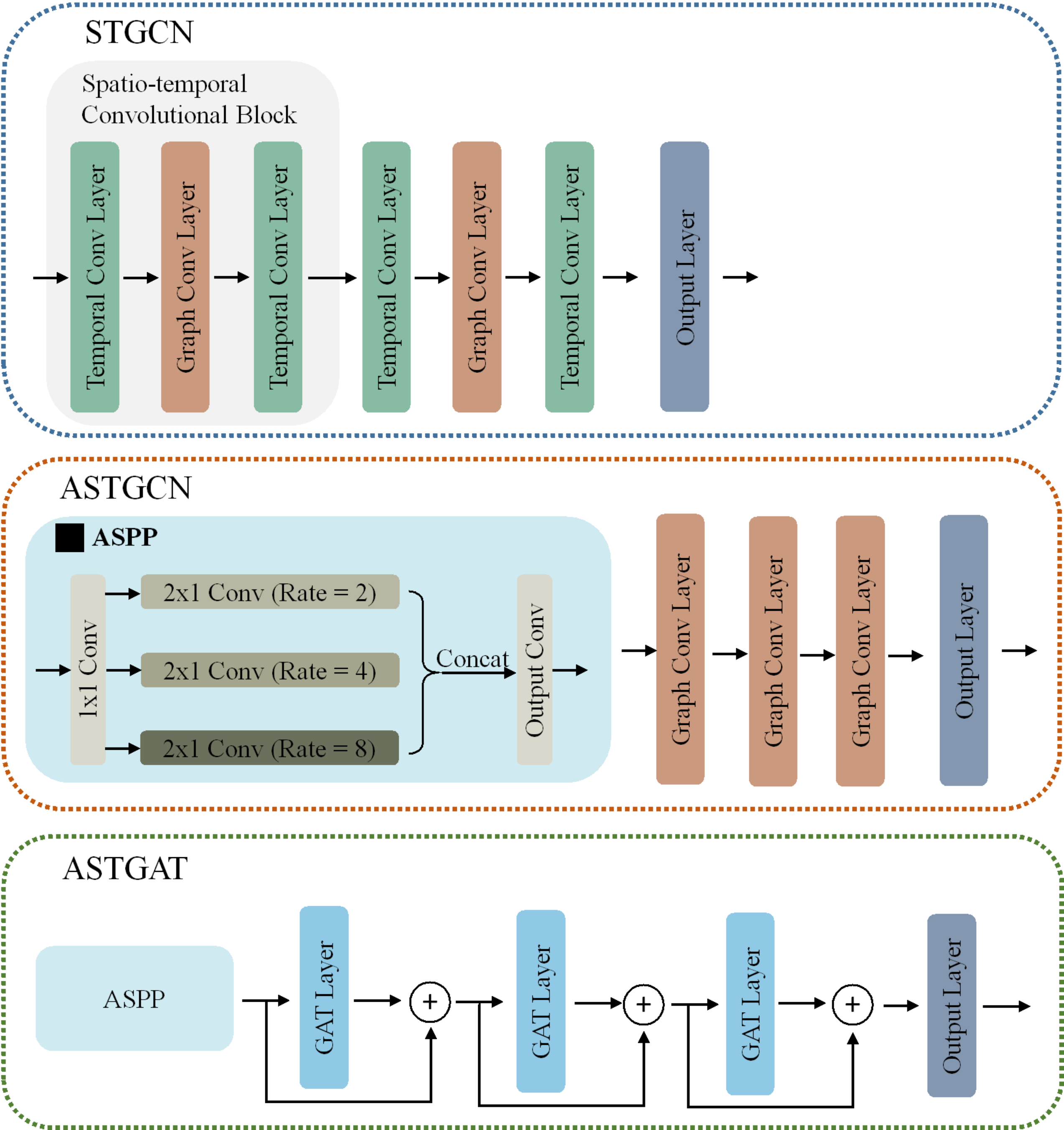}
\caption{The architectures of STGNNs.}
\label{STGNNs}
\end{figure}

Fig. \ref{STGNNs} illustrates the ASPP's architecture (see \ding{110}), which first utilizes a 1x1 convolutional layer followed by three 2x1 convolutional layers, where the dilation rate is gradually increased to capture temporal features at various intervals. After concatenating the multi-scale temporal features, an output convolutional layer is employed to convert the dimension of the temporal feature vectors into the desired size.

\subsection{Models}
We utilize the components described above to construct three distinct STGNNs, as shown in Fig. \ref{STGNNs}, which include STGCN, ASTGCN, and ASTGAT. For STGCN, we adopt the same architecture as in the work \cite{STGCN_TCAD}. The "sandwich" structure is employed to create a spatio-temporal convolutional block. Layer normalization \cite{LayerNorm} is used in the end of the block. Importantly, the normalization layer present in the STGCN architecture is omitted in Fig. \ref{STGNNs}, as well as in the two remaining STGNNs. The output layer converts the detailed features into final predictions, maintaining the same structure as in our previous work.  

In ASTGCN, we use ASPP to capture the temporal feature and three stacked Graph Convolutional Layer described in Section \ref{GCL} to learn the spatial feature. The same normalization layer as STGCN is employed after the input and output layer of ASPP, and after each Graph Convolutional Layer. For ASTGAT, we use the Graph Attention Layer to replace the Graph Convolutional Layer in ASTGCN. In addition, residual connection \cite{ResNet} is employed to prevent vanishing gradients and degradation, as shown in Fig. \ref{STGNNs}.

\section{Experiments}
This section presents the experimental evaluation of the three STGNN models, comparing their performance on the datasets in Table \ref{tab:datasets} to guide future applications. The feasibility of applying the approach to new test cases within a circuit is also examined. The STGNN platform is implemented using PyTorch Geometric \cite{PyG}, an official PyTorch library for developing and training GNNs. This differs from the work \cite{STGCN_TCAD}, which employed the independently developed Deep Graph Library (DGL) \cite{DGL}. Cross Entropy \cite{CE} is used as the loss function during training.

\subsection{Experiment pipeline}
Hybrid sampling is employed to partition SEU samples from each circuit into training, validation, and test sets. In Section \ref{experimentsOnSTGNNs}, 60\% of the samples are used for training, 20\% for validation, and the remaining 20\% for testing. In Section \ref{experimentsOnTestcases}, 75\% of the Ibex samples in Table \ref{tab:datasets} are used for training and 25\% for validation, and test data are generated from two additional Ibex test cases to evaluate model performance.

Two hyperparameters are tuned for the data fed into STGNNs: $max\_distance$ and $time\_win\_size$. The $max\_distance$ controls graph connectivity and determines the number of edges. A small value may yield insufficient structural information, while a large value may introduce noise and increase training complexity. The optimal $max\_distance$ is selected from the range 6 to 10 for each circuit. The $time\_win\_size$ defines the temporal window for each SEU injection time point. Similarly, it should balance capturing adequate sequential features without adding noise. Candidate values of 20, 30, 40, 50, and 60 are evaluated to find the best setting.

For each circuit and STGNN model, the optimal values of $max\_distance$ and $time\_win\_size$ are selected based on validation accuracy, while the final performance is reported on test datasets. To account for the inherent randomness in ML training, the complete pipeline, including training, hyperparameter tuning, and testing, is repeated with different random seeds and hardware (Intel Xeon E5-4627V2 CPU or NVIDIA H100 GPU). Standard evaluation metrics (precision, recall, and accuracy) are computed on the test dataset, and their mean and standard deviation are reported to evaluate performance.

\renewcommand{\arraystretch}{1.2}
\begin{table}[t]
     \setlength\tabcolsep{1.5pt} 
	\centering
	\fontsize{8}{10}\selectfont
	\begin{threeparttable}
		\caption{Performance of STGNNs on test circuits. For example, the accuracy of ASTGAT on the SPI circuit is $96.15 \pm 1.27$, indicating a mean accuracy of 96.15\% with a standard deviation of 1.27\%. }
		\label{tab:resultsOnCircuits}
		\begin{tabular}{ccccccccccc}
			\toprule[1.1pt]
			\multirow{2}{*}{Circuit}&
			\multirowcell{2}{STGNN \\Model}&
			\multirowcell{2}{Precision \\ (\%)}&
			\multirowcell{2}{Recall \\ (\%)}&
			\multirowcell{2}{Accuracy \\ (\%)}\cr\cr 
			\midrule
			\multirow{3}{*}{SPI}&STGCN&$92.29 \pm 2.92$&$91.54 \pm 1.80$&$92.49 \pm 2.19$\cr
			&ASTGCN&$95.80 \pm 0.74$&$93.76 \pm 1.51$&$95.12 \pm 1.04$\cr
			&ASTGAT&$96.17 \pm 0.98$&$95.58 \pm 1.88$&$\mathbf{96.15 \pm 1.27}$\cr
			\midrule
			\multirow{3}{*}{I2C}&STGCN&$94.86 \pm 0.59$&$94.20 \pm 0.90$&$\mathbf{95.50 \pm 0.66}$\cr
			&ASTGCN&$92.96 \pm 0.13$&$91.80 \pm 0.56$&$93.77 \pm 0.30$\cr
			&ASTGAT&$93.24 \pm 0.60$&$92.67 \pm 1.21$&$94.21 \pm 0.77$\cr
			\midrule
			\multirow{3}{*}{XGE-MAC}&STGCN&$93.06 \pm 2.82$&$93.08 \pm 2.80$&$93.01 \pm 2.88$\cr
			&ASTGCN&$92.08 \pm 2.64$&$92.10 \pm 2.61$&$92.08 \pm 2.66$\cr
			&ASTGAT&$93.77 \pm 1.82$&$93.85 \pm 1.82$&$\mathbf{93.80 \pm 1.84}$\cr
			\midrule
			\multirow{3}{*}{ETH-MAC}&STGCN&$97.50 \pm 1.42$&$96.11 \pm 1.64$&$\mathbf{98.90 \pm 0.50}$\cr
			&ASTGCN&$98.06 \pm 0.57$&$93.45 \pm 2.75$&$98.54 \pm 0.53$\cr
			&ASTGAT&$58.60 \pm 23.23$&$61.95 \pm 20.69$&$92.63 \pm 3.92$\cr
			\midrule
			\multirow{3}{*}{Ibex}&STGCN&$90.09 \pm 1.32$&$84.84 \pm 1.59$&$93.26 \pm 0.81$\cr
			&ASTGCN&$89.26 \pm 0.14$&$84.64 \pm 2.21$&$92.94 \pm 0.99$\cr
			&ASTGAT&$90.91 \pm 2.20$&$87.25 \pm 2.63$&$\mathbf{94.00 \pm 0.90}$\cr
			\midrule
			\multirow{3}{*}{RI5CY}&STGCN&$95.24 \pm 0.47$&$93.60 \pm 0.71$&$\mathbf{95.73 \pm 0.37}$\cr
			&ASTGCN&$94.23 \pm 0.34$&$92.57 \pm 0.49$&$94.97 \pm 0.29$\cr
			&ASTGAT&$94.00 \pm 0.71$&$93.19 \pm 0.32$&$95.09 \pm 0.38$\cr
			\hline
			\bottomrule[1.5pt]
		\end{tabular}
	\end{threeparttable}
\end{table}
\vspace{-0.2cm}

\subsection{Experiments on different STGNNs}
\label{experimentsOnSTGNNs}
The performance of STGCN, ASTGCN, and ASTGAT on each circuit is summarized in Table \ref{tab:resultsOnCircuits}. For each STGNN and circuit, the experimental pipeline is executed four times. As shown in the results, the best-performing STGNN model varies by circuit. STGCN and ASTGAT generally outperform ASTGCN, indicating their suitability for new datasets. The best accuracy that each circuit achieves is over 93\%, with precision above 90\% and recall above 87\%, regardless of circuit complexity (e.g., Ibex, RI5CY). ASTGCN demonstrates the most stable performance across circuits. For example, STGCN's accuracy is approximately 4\% lower than that of ASTGAT on SPI, while ASTGAT's accuracy is about 6\% lower than STGCN's on ETH-MAC. In contrast, ASTGCN consistently achieves accuracy relatively close to the highest accuracy obtained on the circuit.

\subsection{Experiments across different test cases}
\label{experimentsOnTestcases}

The work \cite{STGCN_TCAD} proposed a method to accelerate SEU fault simulation, where the ML pipeline, comprising dataset generation, training, tuning, and prediction, was applied to a single test case of a target circuit. To evaluate the generalization capability of the approach, we evaluate the feasibility of applying a model trained on one test case to predict SEU fault simulation outcomes for other test cases on the same circuit. Experiments are conducted on the Ibex, which includes a UVM testbench with diverse functional verification cases. Two additional test cases are selected for evaluation, each involving 20 SEU injection points, resulting in 42,520 SEU samples per case ($2126 \times 20$).

The experiment pipeline is repeated five times, and the results on new test cases are summarized in Table \ref{tab:resultsOnTC}. As expected, accuracy is lower than in Table \ref{tab:resultsOnCircuits}, likely due to different temporal behaviors across test cases, which could pose challenges for STGNN's predictions. Nevertheless, the achieved accuracy remains acceptable (approximately 89\%), demonstrating the potential of the STGNN approach to accelerate SEU fault simulation in scenarios involving multiple test cases or workloads for reliability analysis.

\renewcommand{\arraystretch}{1.2}
\begin{table}[t]
     \setlength\tabcolsep{4pt} 
	\centering
	\fontsize{8}{10}\selectfont
	\begin{threeparttable}
		\caption{Performance of STGNNs on new test cases of Ibex.}
		\label{tab:resultsOnTC}
		\begin{tabular}{ccccccccccc}
			\toprule[1.1pt]
			\multirow{2}{*}{Test case}&
			\multirowcell{2}{STGNN \\model}&
			\multirowcell{2}{Precision \\ (\%)}&
			\multirowcell{2}{Recall \\ (\%)}&
			\multirowcell{2}{Accuracy \\ (\%)}\cr\cr 
			\midrule
			\multirow{3}{*}{test2}&STGCN&$52.90 \pm 2.99$&$54.57 \pm 3.27$&$\mathbf{89.19 \pm 0.71}$\cr
			&ASPPGCN&$41.95 \pm 4.14$&$62.21 \pm 6.52$&$85.71 \pm 1.49$\cr
			&ASPPGAT&$44.24 \pm 3.15$&$60.60 \pm 4.22$&$86.61 \pm 1.09$\cr
			\midrule
			\multirow{3}{*}{test3}&STGCN&$61.83 \pm 4.74$&$57.31 \pm 4.11$&$\mathbf{88.29 \pm 1.26}$\cr
			&ASPPGCN&$54.97 \pm 4.41$&$60.15 \pm 6.99$&$86.65 \pm 1.33$\cr
			&ASPPGAT&$62.71 \pm 7.55$&$58.51 \pm 5.86$&$\mathbf{88.28 \pm 1.93}$\cr
			\hline
			\bottomrule[1.5pt]
		\end{tabular}
	\end{threeparttable}
\end{table}
\vspace{-0.2cm}

\begin{figure}[h]
 \centering
      \subfloat[STGCN] { \label{GenOnSTGCN}
	   \includegraphics[width=0.45\textwidth]{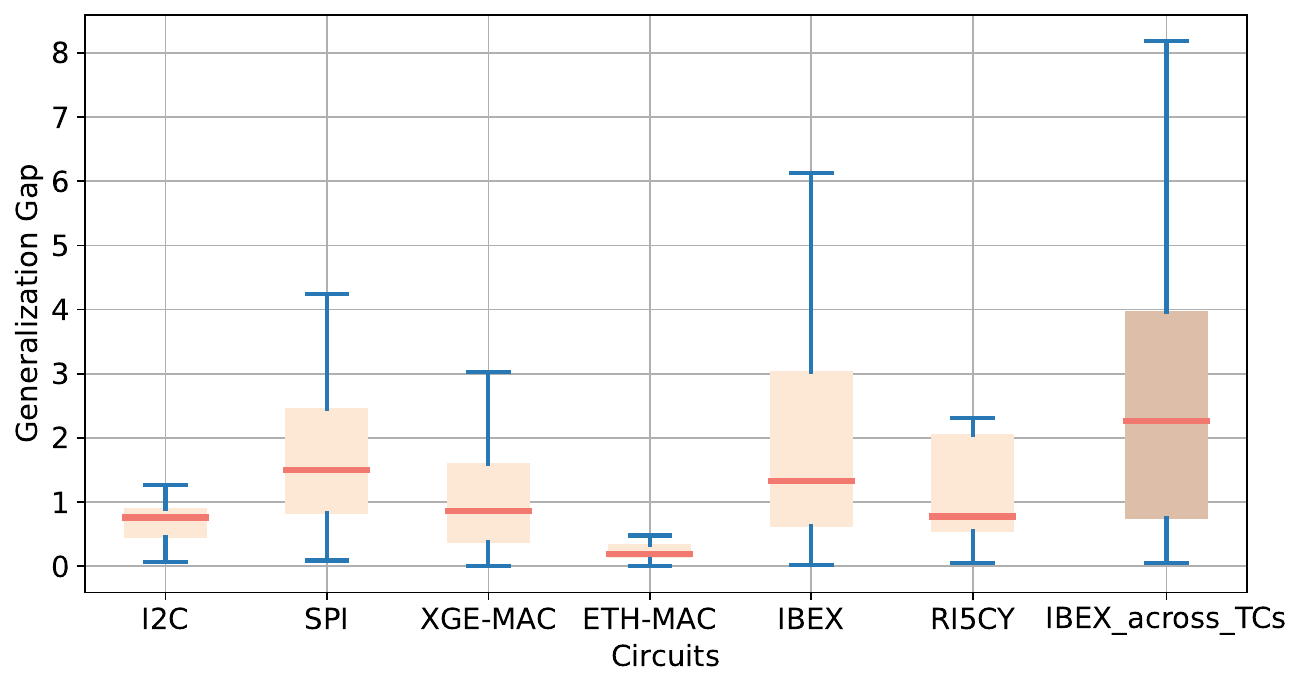} }
       \newline
	 \subfloat[ASTGCN] { \label{GenOnASTGCN}
	   \includegraphics[width=0.45\textwidth]{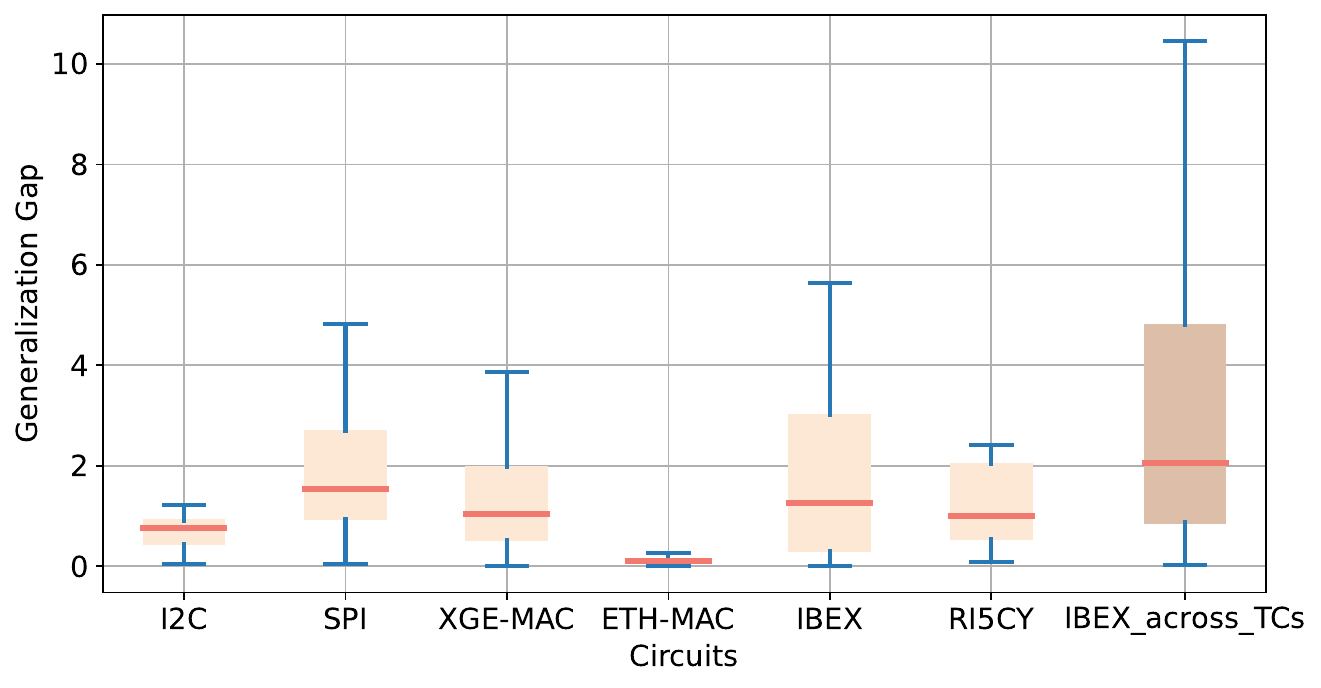} }
      \newline
      \subfloat[ASTGAT] { \label{GenOnASTGCN}
	   \includegraphics[width=0.45\textwidth]{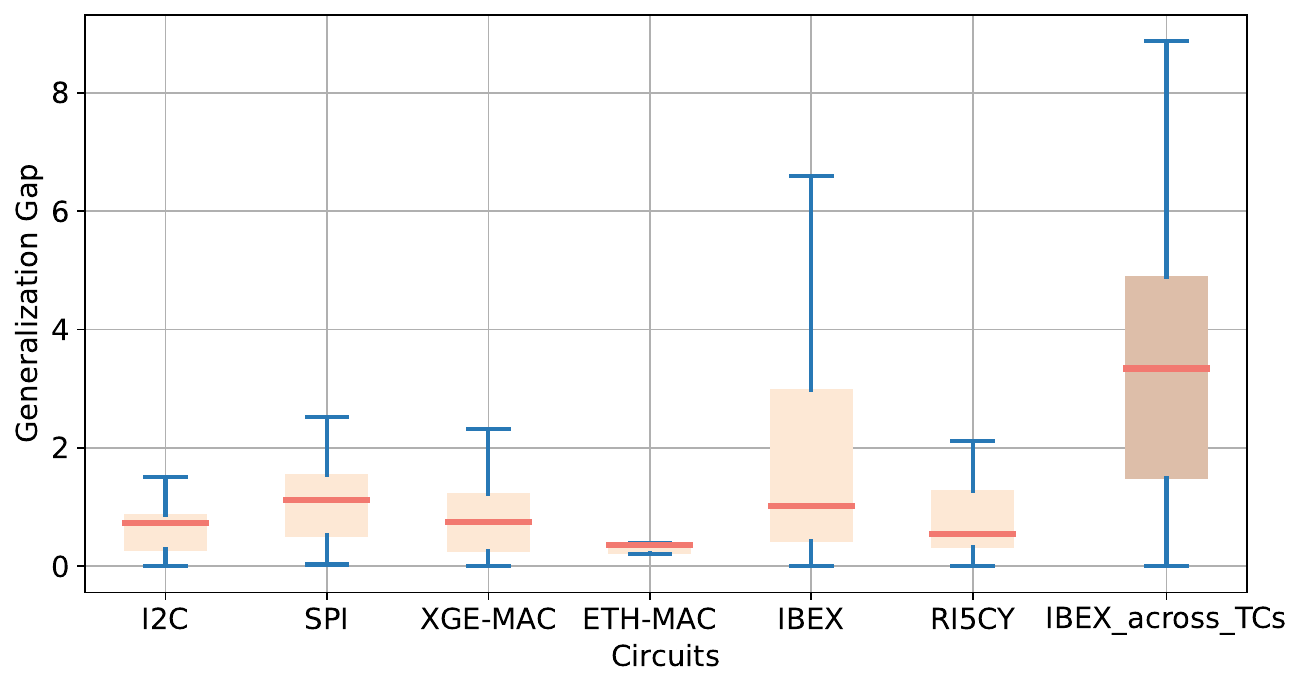} }	
	\caption{The generalization gap in accuracy on validation and test datasets. The box extends from the 25th percentile to the 75th percentile, and lines extend from the box to the smallest and largest values. Wider boxes or longer lines indicate greater variability in generalization. The red line inside the box represents the median of the data.}
	\label{testDiff}
\end{figure}

\subsection{Generalization capability analysis}
When applying the STGNN approach to a practical problem to accelerate SEU fault simulation, the prediction accuracy on test datasets is unknown. To calculate the total accuracy, a commonly used method is to use the accuracy of validation datasets to estimate the prediction accuracy of test datasets. In this context, the generalization capability of ML models is crucial. If the accuracy of a test dataset significantly differs from that of the validation dataset, there may be considerable disparities between predicted and actual conditions. 

To evaluate the generalization capability of the STGNN models, we compute the difference between validation and test accuracy for each experiment, including that conducted during hyperparameter tuning. The distribution of generalization gaps for each circuit is illustrated in Fig. \ref{testDiff}. As shown, the gap between validation and test accuracy remains within 3\% in most cases for experiments performed on the same test case (Section \ref{experimentsOnSTGNNs}), indicating that validation accuracy serves as a reliable estimator of test performance. The most significant generalization gap is observed in the Ibex circuit, likely due to the complex temporal features present in its datasets, which are generated by a UVM-based test bench executing random instructions. For all three STGNN models, the generalization capability is consistently lower when predictions are conducted on different test cases (see \textit{IBEX\_across\_TCs} in Fig. \ref{testDiff}). However, the difference remains below 5\% in most cases.

\section{Conclusions}
This paper presents an open-source platform for predicting SEU fault simulation outcomes using STGNNs. It includes implementations of three STGNN models, along with datasets from six open-source circuits. Model performance is evaluated across the datasets to guide future applications. Furthermore, the feasibility of using trained STGNNs to predict SEU simulation results for unseen test cases within the same circuit is investigated. Although accuracy decreases in cross-test scenarios, the approach remains effective for accelerating reliability analysis across multiple workloads.

\bibliographystyle{IEEEtran}
\bibliography{myRef}

\end{document}